\documentclass[nohyper,11pt,letterpaper]{JHEP3}
\usepackage[dvips]{epsfig}
\usepackage{epsf,amsfonts,amssymb}

\newcommand{\eqn}[1]{(\ref{#1})}
\newcommand{\be}{\begin{equation}}
\newcommand{\ee}{\end{equation}}
\newcommand{\ben}{\begin{displaymath}}
\newcommand{\een}{\end{displaymath}}
\newcommand{\bea}{\begin{eqnarray}}
\newcommand{\eea}{\end{eqnarray}}
\newcommand{\bean}{\begin{eqnarray*}}
\newcommand{\eean}{\end{eqnarray*}}
\newcommand{\nn}{\nonumber \\}
\newcommand{\ba}{\begin{array}}
\newcommand{\ea}{\end{array}}
\newcommand{\bi}{\begin{itemize}}
\newcommand{\ei}{\end{itemize}}

\def\l {\lambda}
\def\a {\alpha}
\def\ap {\alpha'}
\def\b {\beta}
\def\g {\gamma}
\def\G {\Gamma}

\def\s {\sigma}
\def\e {\epsilon}

\def\t {\theta}
\def\rl {\sqrt{\lambda}}
\def\undos{{1\over 2}}

\newcommand{\bref}[1]{(\ref{#1})}

\newcommand{\call}{\mbox{${\cal L}$}}

\newcommand{\caln}{\mbox{${\cal N}$}}
\newcommand{\calo}{\mbox{${\cal O}$}}

\newcommand{\bbr}[1]{\mbox{${\mathbb R}^{#1}$}}

\newcommand{\bbi}[1]{\mbox{${\mathbb I}_{#1}$}}

\newcommand{\ads}[1]{\mbox{${AdS}_{#1}$}}
\newcommand{\adss}[2]{\mbox{$AdS_{#1}\times {S}^{#2}$}}

\newcommand{\fc}{\frac}
\newcommand{\w}{\omega}

\newcommand{\sac}{\ , \qquad}

\newcommand{\ie}{{\it i.e.}}

\newcommand{\ra}{\rightarrow}

\newcommand{\sect}[1]{\setcounter{equation}{0}\section{#1}}

\newcommand{\beq}{\begin{equation}}
\newcommand{\eeq}{\end{equation}}
\newcommand{\beqr}{\begin{displaymath}}
\newcommand{\eeqr}{\end{displaymath}}
\newcommand{\beqa}{\begin{eqnarray}}
\newcommand{\eeqa}{\end{eqnarray}}
\newcommand{\beqar}{\begin{eqnarray*}}
\newcommand{\eeqar}{\end{eqnarray*}}

\title{\LARGE Supersymmetry of Tensionless Rotating Strings
              in $AdS_5\times S^5$, and Nearly-BPS Operators}

\author{David Mateos \\
  Perimeter Institute for Theoretical Physics \\
  Waterloo, Ontario N2J 2W9, Canada \\
  \email{dmateos@perimeterinstitute.ca}}

\author{Toni Mateos \\
  Departament ECM, Facultat de F\'\i sica \\
  Universitat de Barcelona, Institut de F\'\i sica
  d'Altes Energies and \\
  CER for Astrophysics, Particle Physics and Cosmology \\
  Diagonal 647, E-08028 Barcelona, Spain \\
  \email{tonim@ecm.ub.es}}

\author{Paul K.\ Townsend \\
  Department of Applied Mathematics and Theoretical Physics \\
  Centre for Mathematical Sciences \\
  Wilberforce Road, Cambridge CB3 0WA, United Kingdom \\
  \email{p.k.townsend@damtp.cam.ac.uk}}

\abstract{It is shown that a class of rotating strings in \adss{5}{5}
with $SO(6)$ angular momenta $(J,J',J')$ preserve 1/8-supersymmetry
for large $J,J'$, in which limit they are effectively tensionless;
when $J=0$, supersymmetry is enhanced to 1/4.
These results imply that recent checks of the AdS/CFT
correspondence actually test a nearly-BPS sector.}

\keywords{D-branes, Supersymmetry and Duality}

\preprint{\tt{ECM-2003-24}  \\ \tt{DAMTP-2003-78} \\ \tt{hep-th/0309114} }

\begin{document}

\section{Introduction} \label{intro}

The AdS/CFT correspondence states that $\caln =4$ super Yang-Mills 
(SYM) theory is the holographic dual of IIB string theory on \adss{5}{5}
\cite{Maldacena97, GKP98, Witten98}. The latter is usually said to reduce 
to
supergravity in a weak-coupling and infinite-tension limit. However, 
this limit also validates a semi-classical quantization of macroscopic 
strings \cite{GKP02}. Examples of these that will be of interest here are
provided by closed strings on $S^5$ that lie at the origin of \ads{5} and
are supported against collapse by their $SO(6)$ angular momenta
\cite{FT03a,FT03b,FT03c}.

Closed strings in flat space that are supported by angular momentum were 
first found by Hoppe and Nicolai \cite{hoppe}, and their stability 
established in \cite{stable}. These Hoppe-Nicolai configurations  actually
lie on a sphere, so it is not surprising that they also solve the
equations of motion when flat space is replaced by \adss{5}{5}, as was
discovered by Frolov and Tseytlin (FT) \cite{FT03a}, who also found a
more general type of configuration supported by two independent angular
momenta.  An analysis of the stability of these rotating string solutions
has revealed that they are typically unstable.\footnote{
As the quadratic fluctuations depend on the curvature of the
ambient space, the stability properties in \adss{5}{5} can be quite 
different from those in flat space.} Despite this, spectacular agreement
has been found between their energies and the anomalous dimensions of the
corresponding CFT dual  operators, in the limit of large angular momenta 
\cite{BMSZ03, FT03c, AFRT03}. It has been suggested that 
this is a test of the AdS/CFT correspondence in a `far-from-BPS
sector'. 

It is true that the FT rotating strings break all supersymmetries, 
but an inspection of the stability results of these authors
shows that any instability disappears in the limit of infinite angular
momentum. Furthermore (as will be shown) a BPS-type bound on the energy is
saturated in this limit. This suggests that the FT rotating string
solutions might become supersymmetric in this limit, in which case the
test of the AdS/CFT correspondence that they provide would actually be in
a `nearly-BPS' sector. We say `nearly-BPS' instead of `near-BPS' because
the limit of infinite angular momentum is also a limit of infinite energy
and cannot actually be taken; there is no supersymmetric rotating string
solution `near to' the solutions of Frolov and Tseytlin. Nevertheless,
the infinite-energy limit is an ultra-relativistic limit in which the
string is effectively tensionless and, as we will show, this tensionless
string {\it is} supersymmetric. Since an FT string at large angular
momentum is `nearly tensionless' it is also `nearly supersymmetric'.

We establish this result using the by now standard procedure involving 
the `$\kappa$-symmetry' transformations of the IIB Green-Schwarz 
superstring fermions. This computation reveals that a tensionless FT
string preserves 1/4 or 1/8 of the supersymmetry, according to whether
the rank of the angular-momentum two-form is four or six, respectively.
Because \adss{5}{5} is conformally flat, it might appear
that there is no difference between \adss{5}{5} and flat space for 
a tensionless string. However, \adss{5}{5} is not 
{\it super}conformally flat \cite{superconformal},  and indeed the 
supersymmetric configurations that we find in \adss{5}{5} are {\it not} 
supersymmetric in flat space.

\sect{Rotating Strings}
\label{rot} 

We will be considering solutions of the classical equations of motion 
derived
from the Nambu-Goto Lagrangean
\be
\call = - \fc{1}{\ap} \,  \sqrt{-\det g} \,,
\ee
where $g$ is the worldsheet metric induced from the \adss{5}{5} spacetime
metric. We begin with a review of the relevant rotating string solutions 
of \cite{FT03a}, which are strings that lie at the origin of \ads{5} and
rotate only in the $S^5$, so effectively they live in a 
$\bbr{} \times S^5$ subspace of \adss{5}{5} with metric 
\be 
ds^2 = R^2 \, \left( - dt^2 + d\Omega_{\it 5}^2 \right) \,, 
\ee 
where the constant $R$ is related to the inverse string tension, 
$\alpha'$, and the 't Hooft coupling, $\lambda$, by 
$R^2 = \ap \sqrt{\lambda}$, and $d\Omega_{\it 5}^2$ is the 
$SO(6)$-invariant metric on the unit five-sphere.
The latter may be viewed as the submanifold $|{\bf W}|=1$ of $\bbr{6}$
with Cartesian coordinates $W_i$ ($i=1,\dots,6$). For the
parametrization with 
\bea
W_1 + i W_2 &=& \cos \t \, e^{i \chi} \,, \nn 
W_3 + i W_4 &=& \sin \t \, \cos \phi
\, e^{i \a} \,, \nn 
W_5 + i W_6 &=& \sin \t \, \sin \phi \, e^{i \b} \,, 
\label{y} 
\eea 
this gives 
\be 
d\Omega_{\it 5}^2 = d\t^2 +
\sin^2 \t \, d\phi^2 + \cos^2 \t \, d\chi^2 + \sin^2 \t \cos^2 \phi 
\, d\a^2 +
\sin^2 \t \sin^2 \phi \, d\beta^2  \,. 
\label{metric} 
\ee

We fix the worldvolume reparametrization invariance by the gauge
choice
\be
t = \tau \sac \a=\s \,,
\ee
where $\tau$ and $\s$ are the worldsheet coordinates.\footnote{
  Note that in our conventions both the spacetime and
  the worldsheet coordinates are dimensionless, which implies that 
  the energy of the string, $E$, is also dimensionless. The 
  corresponding dimensionful time and energy are obtained by rescaling 
  the dimensionless ones by appropriate powers of $R$.} 
The string solutions of interest correspond to circular, rotating 
strings supported against collapse by their angular momenta; 
they are given by 
\be 
\t = \t_0 \sac \chi = \nu \tau \sac \phi = \omega \tau \sac \b = \s \,,
\label{embedding} 
\ee 
where $\t_0$ is a constant in the interval $[0,\pi/2]$.
At any instant the string is a circle in a two-plane contained within the
3456-space; this plane rotates with angular velocity $\omega$. In turn, 
the string's centre of mass rotates with angular velocity $\nu$ around a
circle in the 12-plane.

Under the above circumstances, the Nambu-Goto equations are solved if 
either of the following relations hold\footnote{Note that in case $(i)$
the string always lies at the origin of the 12-plane, which is a singular
submanifold of the coordinate system we are using; $\nu$ is the angular
velocity in a circle of zero radius. In practice this does not cause
a problem because $\nu$ is irrelevant in this case.}
\bea 
(i)&& \qquad \cos \t_0 = 0 \,, \qquad \w^2<1 \,, \nn 
(ii)&& \qquad \cos 2\t_0 = {\omega^2 -1 \over \omega^2 -\nu^2} \,, 
\qquad \nu^2<1 \,, \qquad 2 \w^2 - \nu^2-1>0 \,.
\eea 
The restrictions on the angular velocities follow from demanding the
reality of both $\call$ and $\t_0$. 
The energy of the rotating string is
\be
E = \rl \, |\sin \t_0| \, \Delta^{-1/2} \,, \qquad \Delta \,\equiv \, 1
-\nu^2 \cos^2 \t_0  -\w^2 \sin^2\t_0  \,, 
\label{e}
\ee
while the only non-zero components of the angular momentum two-form, 
$J_{ij}$, in the Cartesian coordinates $W_i$, are
\be 
J \equiv J_{12} = E\, \nu \, \cos^2 \t_0 \sac
 J' \equiv J_{35} = J_{46} = \fc{1}{2} \, E \, \omega \, \sin^2 \t_0 \,. 
\label{j} 
\ee 
Thus, $J$ and $2J'$ are the momenta conjugate to $\chi$ and $\phi$,
respectively.
Their values for the two possible solutions
of the equations of motion are
\bea 
(i) && E = \fc{\rl}{\sqrt{1 - \omega^2}} 
            \sac J = 0  \sac  J' = \fc{\rl \, 
\omega}{\sqrt{1- \omega^2}} \,, 
\\
(ii)     && E = \fc{\rl}{\sqrt{\omega^2 - \nu^2}} \sac
 J = \rl \, \fc{\left(2\omega^2 -\nu^2-1\right)\, \nu} 
{2\left(\omega^2 -\nu^2\right)^{3/2}} \sac  J' = \rl \,\, 
\fc{\omega\left(1-\nu^2\right)} 
{2\left(\omega^2 -\nu^2\right)^{3/2}} 
 \,. \nonumber 
\eea 
In the first case it is easy to express 
the energy solely as a function of the angular momentum, with the result 
\be
(i) \qquad E = \sqrt{(2J')^2 + \l} = 
|2J'| \, \left[ 1 + \calo \left(\fc{\l}{J'^2} \right) \right] \,.
\ee
The expression for $E$ in terms of $J$ and $J'$ can also be found 
explicitly in the second case, but the result is rather messy. However, 
when
expanded for large $J, J'$, it yields 
\be 
(ii) \qquad E = \left(|J| + |2J'|\right) \, \left[ 1 + 
\calo \left(\fc{\l}{J^2}, \fc{\l}{J'^2} \right) \right]\,.
\ee
In either case, the leading, $\lambda$-independent terms saturate the BPS
bounds on the energy that we will derive from the \adss{5}{5} superalgebra in
the following section. Thus the bounds are asymptotically saturated in
the limit of large angular momenta. This  result suggests that the circular,
rotating strings become supersymmetric in this limit; we will confirm 
this by an explicit calculation in Section \ref{susy}.

At first sight, our results seem to conflict with the analysis of 
perturbative stability of these solutions in \cite{FT03a,FT03b}. 
It was shown there that the solutions $(i)$ are unstable 
if $J > 3 \rl /8$, and that those of $(ii)$ are only stable if 
$J>2J'/3$. These instabilities are related to some fluctuations about 
the solutions having tachyonic masses.\footnote{
The relevant tachyonic masses appear in formulas (4.34) of
\cite{FT03a} and (2.35), (2.36) of \cite{FT03b}.} 
However, these masses scale as $m^2 \sim - E^{-2}$, where $E$ is the
energy of the classical solution. In the large angular momenta 
limit, which is also a large energy limit, these masses tend to zero,
so that all instabilities disappear asymptotically.

\sect{BPS Bound from the Superalgebra}
\label{section-bound} 

The energy of supersymmetric string states in \adss{5}{5} must saturate 
a  BPS bound that follows from the $PSU(2,2|4)$ isometry superalgebra of
the \adss{5}{5} vacuum, and hence their energy can be expressed as a
function of their charges alone. In this section we review the BPS bound
for states that carry the same type of charges as the rotating, circular
strings above, that is, energy and angular momenta on the $S^5$. The BPS
bound we will derive may be equally well understood as a statement about
the supersymmetry properties of operators in the dual CFT; we will
come back to this point in the penultimate section.

Let $\gamma_m$ ($m=0,\ldots,4$) be the $4\times 4$ five-dimensional 
Dirac matrices for $AdS_5$ and let $Q^i$ be the four $AdS_5$ Dirac 
spinor charges, transforming as the {\bf 4} of $SU(4)$. 
The non-zero anticommutators are
\be\label{5susy}
\{Q^i,Q^\dagger_j\} = \gamma^0 \left[
\left(\gamma_m P^m + {1\over2}\gamma_{mn}M^{mn}\right)\delta^i{}_j
+ 2 \bbi{}\, B^i{}_j  \right] \,,
\ee
where $P, M$ are the $AdS_5$ charges and $B$ is the hermitian
traceless matrix of $SU(4)$  charges. For our spinning string
configurations the only non-zero $AdS$ charge  is the energy $P^0 =E$; in
this case \bref{5susy} reduces to 
\be
\{Q^i,Q^\dagger_j\} =\bbi{}\, \delta^i{}_j  E + 2\gamma^0 B^i{}_j \,.
\ee
By means of an $SU(4)$ transformation we may bring $B$ to diagonal form
with diagonal entries $b_i$ ($i=1,2,3,4$) satisfying
\be 
\label{trace}
b_1+b_2+b_3+b_4 =0.
\ee
The eigenvalues of the $16\times 16$ matrix $\{Q,Q^\dagger\}$ are
therefore 
$E\pm 2b_1, E\pm b_2, E\pm b_3,E\pm b_4$, each being doubly degenerate.
Since this matrix is manifestly positive in any unitary representation,
unitary implies the bound
\be\label{Ebound}
E \ge 2b \qquad b=\mbox{sup}\{ |b_1|,|b_2|,|b_3|,|b_4| \} \,.
\ee
When the bound is saturated the matrix $\{Q,Q^\dagger\}$ will have zero
eigenvalues; the possible multiplicities are $2,4,8,16$. The maximum
number (16) occurs when $b_i=b$ for all $i$, in which case (\ref{trace})
implies $b=0$ and hence $E=0$; this is the $adS_5$ vacuum. Otherwise, one
has preservation of $1/8,1/4,1/2$ supersymmetry when
$\{Q,Q^\dagger\}$ has $2,4,8$ zero eigenvalues, respectively. 

The eigenvalues of $B$ are $SU(4)$ invariants and hence determined in
any $SU(4)$ irrep by that irrep's Dynkin labels $(d_1,d_2,d_3)$.
Conversely, the Dynkin labels are determined by the eigenvalues of $B$,
and consideration of the highest weight state leads to the relation 
\be
d_1 = b_1-b_2, \qquad d_2=b_2 -b_3,\qquad d_3= b_3-b_4.
\ee
Given the constraint \bref{trace}, this can be inverted to give
\bea
b_1 &=& {1\over 4}\left( 3d_1+2d_2+d_3 \right) \,,\nn b_2
&=& {1\over 4}\left( -d_1+2d_2+d_3 \right) \,, \nn 
b_3 &=& {1\over 4}\left( -d_1-2d_2+d_3 \right)\,, \nn 
b_4 &=& {1\over 4}\left( -d_1-2d_2-3d_3 \right) \,.
\label{bvalues}
\eea 
The $SU(4)$ charges of the spinning strings considered here 
correspond to irreps with Dynkin labels \cite{FT03a}
\bea
\label{choiceone}
&& [d_1, d_2, d_3]=[J'-J,0, J + J'] \qquad  \mbox{if}  
\qquad J' > J  \,, \\ \label{choicetwo} 
&&[d_1, d_2, d_3]=[0,J-J',2 J'] \qquad \,\,\,\,\,\,\,\,
\mbox{if} \qquad J \le J'  \,. 
\eea 
It follows, for either case, that the four (unordered) eigenvalues of $B$
are
\be \{  J'-\undos J \, , \, \undos J \, , \,
  \undos J \, , \,  - J' - \undos J \} \,.
\label{js}
\ee
Using this in \bref{Ebound}, we deduce that
\be
E \ge |J| + 2|J'|.
\ee
When this bound is saturated the matrix of anticommutators of
supersymmetry charges will have zero eigenvalues, corresponding to the
preservation of some fraction of supersymmetry. Let us determine this
fraction under the assumption that $J'>J$. In this case the Dynkin
labels are given by \bref{choiceone} and hence
\be
b_1 = J' -{1\over2} J, \qquad b_2 = {1\over2} J,\qquad
b_3 = {1\over2} J,\qquad b_4 = -J' - {1\over2} J.
\ee
Generically,  $|b_1| =b$ and all other eigenvalues of $B$ have absolute
value less than $b$ so the supersymmetry fraction is 1/8. 
However, this fraction is enhanced to 1/4 if $J=0$ because
then $|b_1|=|b_4|=b$ with $|b_2|,|b_3| <b$. 

A similar analysis for $J\ge
J'$ again yields the fraction 1/8 generically, with enhancement to 
1/2 if $J'=0$; in this case the string reduces to a point-like string
orbiting the $S^5$ along an equator, as  considered by Berenstein,
Maldacena and Nastase (BMN) \cite{bmn}. Finally, if $J=J'=0$ then 
$b_i=0$ for all $i$, $E=0$,  and all supersymmetries are preserved, as
expected for the \adss{5}{5} vacuum.

\sect{Supersymmetry}
\label{susy} 

The supersymmetries preserved by a IIB string correspond to complex
Killing spinors $\epsilon$ of the background that 
satisfy\footnote{Thus, 
$\Upsilon/\sqrt{-\det g}$ is the matrix $\G_{\kappa}$ appearing in the 
kappa-symmetry transformation of the fermionic variables of the 
Green-Schwarz IIB superstring.} 
\be\label{kappa}
\Upsilon \, \e =
\sqrt{-\det g} \, \e \sac  
\Upsilon = X'{}^M \dot X^N \gamma_{MN}\, K \,,
\ee
where $K$ is the operator of complex conjugation, and $\gamma_M$ are 
the (spacetime-dependent) Dirac matrices. 

Recall that we are interested in strings that live in the 
$\bbr{} \times S^5$ submanifold of \adss{5}{5} with metric \eqn{metric}, and
whose embedding is specified by equation \eqn{embedding}. 
Under these circumstances\footnote{
As the radius $R$ cancels in the final result we set $R=1$ in this section.}
\be
\sqrt{-\det g} = \sin\t \sqrt{1 - a^2 - b^2} \sac 
\dot X \cdot \gamma = \G_t + a \, \G_{\chi} + b \, \G_\phi \,,
\ee
where $\G_\t, \G_\phi, \ldots$ are 
ten-dimensional tangent space (\ie constant) Dirac matrices, and
\be \label{aandb}
a= \nu \, \cos \t \sac b = \omega \, \sin \t \,.
\ee
Note that the Lorentzian signature of the induced worldsheet metric
implies that
\be \label{rest}
a^2+b^2 \leq 1 \,.
\ee

The Killing spinors
of \adss{5}{5} restricted to the relevant submanifold take the form 
\be \e =
e^{\fc{t}{2} i \tilde{\G}} \,
     e^{\fc{\t}{2} i \g_* \G_\t} \,
     e^{\fc{\phi}{2} \G_{\t\phi}} \,
     e^{\fc{\chi}{2} i \g \G_{\chi}} \,
     e^{\fc{\a}{2} \G_{\t\a}} \,
     e^{\fc{\b}{2} \G_{\phi\b}} \, \e_0 \,,
\ee
where $\e_0$ is a constant spinor, $\g_* = \G_{\t \phi
\chi \a \b}$, and $\tilde{\G}$ is a constant matrix that commutes with all
other matrices above (its specific form will not be needed).  
In our conventions all these matrices are real. For our configurations 
$\dot X\cdot X'=0$, so the supersymmetry preservation condition
(\ref{kappa}) can be written as
\be 
\label{kappanew}
\left(X'\cdot \gamma\right) \left(\dot X \cdot \gamma \right) \epsilon 
=  - \sqrt{-\det g}\ K \, \epsilon \,.
\ee
This must be satisfied for all $\tau, \sigma$, but it is useful to
first consider $\tau=0$, in which case it reduces to 
\be 
\sin\t \left( \dot X \cdot \gamma \right) \, \G_\a \, K \,
e^{\fc{\t}{2} 
i \g_* \G_\t} \,
e^{\fc{\sigma}{2} \left( \G_{\t\a} +  \G_{\phi \b} \right)} \, \e_0 = 
\sqrt{-\det g} \,
e^{\fc{\t}{2} i \g_* \G_\t} \, e^{\fc{\sigma}{2} \left( \G_{\t\a} +  \G_{\phi
\b} \right)} \, \e_0 \,. \label{tau0} 
\ee 
It can be shown that in order for this equation to be satisfied for all $\s$, 
one must impose
\be
\G_{\t\a\phi\b} \, \e_0 =  \, \e_0 \,, \label{cond1} 
\ee 
in which case the equation becomes 
\be \label{condi2}
\left[ \G_t + \left( \cos \t - \sin \t \, i \g_*
\G_\t \right) \left( a \, \G_{\chi} + b \, \G_\phi \right) \right] \, \G_\a \,
K \, \e_0 = \sqrt{1-a^2 - b^2} \, \e_0 \,. 
\ee 
Equations \bref{cond1} and \bref{condi2}  are equivalent to the two equations
\bea
A\, \e_0 & = & \e_0  \sac A \equiv a \, \cos\t \, \G_{t\chi} + b \, \sin\t
\, i \G_{t\chi\a\b} \label{A} \,, \\
B\, K \e_0 & = & \sqrt{1-a^2 - b^2}\, \e_0 \sac 
B \equiv a \, \sin\t \, i \G_{\phi\b} \,  + b \, \cos\t \,
\G_{\phi\a} \,  \,. \label{B}  
\eea 
Given that $a$ and $b$ are non-zero, it follows from (\ref{A}) that 
\be 
i \G_{\a\b} \, \e_0 = s \, \e_0 \sac
\G_{t\chi} \, \e_0 = s \, \e_0 \,,
\ee
and
\be
a \, \cos\t + s \, b \, \sin\t = \tilde{s} \,,
\ee
where $s$ and $\tilde{s}$ are independent signs. 
The latter relation is compatible
with the restriction \bref{rest} if and only if
$b\, \cos\t = s \, a \, \sin\t$, and these two relations
for $a$ and $b$ imply, given \bref{aandb}, that 
\be 
\nu = \tilde{s} \sac \omega = s \, \tilde{s} \,. 
\ee 
It then follows that the equation \bref{B} is
trivially satisfied, and that $\sqrt{-\det{g}}=0$.
The string worldsheet must therefore be null, 
which is only possible for a tensionless string. 
Although the IIB superstring is not
tensionless, the energy due to the rotation is much greater
than the energy due to the tension in the limit of
large angular momentum. So in this limit the string is
effectively tensionless. 

We continue now by considering only the tensionless string, for which the
supersymmetry preserving condition (\ref{kappanew}) reduces to 
\be \label{nullpre}
\left(\dot X \cdot \gamma\right) \, \epsilon =0\, .
\ee
The analysis of this equation for $\tau=0$ reproduces the results already
obtained from an analysis of (\ref{tau0}), which are summarized by the 
projections
\be
\label{finalproj} 
\G_{\t\a\phi\b} \, \e_0 =  \, \e_0 \sac \G_{t\chi} \, \e_0
= \omega \nu \, \e_0 \sac i \G_{\a\b} \, \e_0 = \omega \nu \, \e_0 \,.
\ee 
It is now straightforward to check that a spinor $\e_0$ satisfying these
conditions solves \bref{nullpre} for all $\tau$ and $\sigma$.
It thus follows that the generic null FT string
preserves 1/8 of the 32 supersymmetries of the IIB $AdS_5 \times S^5$
vacuum.

We have assumed above that $a$ and $b$ are non-zero. A solution with
$b=0$ has $J'=0$ and corresponds to a  point-like, collapsed string 
moving along a great circle of $S^5$, as considered in \cite{bmn}, 
whereas a solution with $a=0$ has $J=0$. Redoing the analysis it
is easy to see the former preserve 1/2 of the supersymmetry. Similarly,
in the second case one finds that the necessary and sufficient conditions
for preservation of supersymmetry are
\be 
\omega^2=1 \sac \cos\t_0=0,
\ee 
and that the projections on $\epsilon$ are
\be 
\G_{\t\a\phi\b} \, \e_0 \, = \, 
\e_0 \sac i \Gamma_{t\chi\a\b} \e_0 = \omega\e_0 \,. 
\ee 
These projections preserve 1/4 of the thirty-two supersymmetries of the IIB 
$AdS_5 \times S^5$ vacuum.

\sect{Nearly-BPS Operators}
\label{BPSopes}

Macroscopic, rotating strings in \adss{5}{5} with $SO(6)$ angular momenta 
are expected to be dual, at least in the limit of large angular momenta, to
operators of $\caln=4$ SYM theory in appropriate $SO(6)$ representations, with
the energy of each string equal to an eigenvalue of the matrix of anomalous
dimensions. The rotating strings considered here are expected to correspond to
operators transforming in $SO(6)$ representations with Dynkin labels as in
\eqn{choiceone}, \eqn{choicetwo}. The highest-weight operators
in these representations are linear combinations of traces, or
products of traces, of the scalar operators
\be 
X \equiv W_1 + i W_2 \sac Y \equiv W_3 + i W_5 \sac Z = W_4 + i W_6,
\ee 
associated to the non-zero components of the string's angular momenta, 
$J_{12}, J_{35}$ and $J_{46}$.

For $J,J'\ll \sqrt{N}$, the $p$-trace operators are (approximately)
orthogonal to the $q$-trace operators for $p\ne q$. In this 
limit\footnote{The assumption that $J,J'\ll \sqrt{N}$ is compatible 
with our other assumption that $J,J'\gg \sqrt{\l}$, and the two
together imply that the IIB string theory is weakly coupled.} 
an $n$-string state is associated with an $n$-trace operator. 
As we consider a single string, we expect the $SO(6)$ highest-weight 
dual operator to be a single-trace operator of the form 
\cite{FT03a}
\be\label{calop}
\calo (J,J') \, = \, \mbox{Tr} \left(X^J Y^{J'} Z^{J'} \right) 
+ \cdots \,, 
\ee
where the dots stand for the permutations of the factors needed so
that the operator transforms in the appropriate irrep of $SO(6)$.
Evidence for this correspondence is that the anomalous dimensions of 
the $\calo$-type operators have been computed by spin-chain methods 
in the one-loop planar approximation \cite{BMSZ03}, and perfect
agreement has been found with the string prediction in the large 
angular momenta limit \cite{FT03a,FT03b,FT03c}. Note that the spin
chain computation implicitly assumes that $J,J' \ll \sqrt{N}$,
because this condition is needed to justify the restriction to  planar
diagrams; as in the BMN case, non-planar diagrams are expected to be
suppressed by powers of $J^2/N$, $J'^2/N$. 

Our results concerning the supersymmetry of the rotating strings dual to 
the $\calo$-type operators in the limit of large angular momenta imply 
that these operators are `nearly-BPS' in this limit, in a sense that
we now aim to clarify. Given that anomalous dimensions of these
operators are known only for $J,J'\ll \sqrt{N}$, we shall continue to 
assume, for the moment, that the same condition holds on the string
theory side of the correspondence, and we begin by summarising what it
means for an operator to be a BPS operator. All primary operators in a
superconformal multiplet can be obtained  by the action of the
$Q$-supersymmetry charges on a lowest-dimension operator, a so-called
{\it super}conformal primary operator, which commutes with all the
$S$-supersymmetry charges. The superconformal algebra implies that a
superconformal primary operator also commutes with some $Q$-supersymmetry
charges if and only if its conformal dimension saturates an appropriate
BPS bound in terms of its rotational and  R-symmetry quantum numbers
(this is equivalent to the bound on the adS energy that we derived in
section \ref{section-bound}).  If this happens the corresponding
supermultiplet is shorter than a generic supermultiplet, and each
operator in a shortened multiplet is called a BPS operator.  Note that
merely commuting with some Poincar\'e supersymmetry charges does not make
an operator a BPS operator. 

Now we examine the $\calo$-type operators of \bref{Ebound}. These are
primary (after diagonalisation of the matrix of anomalous dimensions)
but not superconformal primary; for example, the operator with
$J=0$, $J'=2$ is a descendant of the Konishi operator. Moreover, they are
not 1/4-BPS or 1/8-BPS operators, since in $\caln =4$ SYM such BPS
operators are linear combinations of multi-trace operators that involve
at least\footnote{In the free theory, $\l=0$, there exist
purely double-trace 1/4-BPS  and purely triple-trace 1/8-BPS operators
\cite{AFSZ99}.} a double-trace or a triple-trace operator, respectively
\cite{AFSZ99,Ryzhov01,DDHR03}. Therefore, although $\calo$ is a
nearly-BPS operator in the sense that its conformal dimension almost
saturates a BPS bound when $\l/J^2 , \l/J'^2 \gg 0$, it is not
the case that $\calo$ approximates an exact 1/4-BPS or 1/8-BPS
operator in this regime. In this sence the $\calo$-type operators are not
`near-BPS', but they are effectively so for any computation that depends
only on the conformal dimension and R-symmetry quantum numbers. We call
these operators `nearly-BPS'. 

To actually take the limit $\l/J^2 , \l/J'^2\ra 0$ we would need to go to
the free theory, $\l=0$. In this case, the conformal dimension of
$\calo$ does saturate a BPS bound, and therefore must belong to a
shortened supermultiplet. This is possible because operators that are
descendents of superconformal primaries in the interacting theory can
become independent BPS operators in the free-field limit
\cite{DO03,HH03}.  Note that there will be as many of these additional
shortened multiplets as are required to form a long one, so the
shortening provides no  protection against the generation of large
anomalous dimensions: the usual claim that BPS-operators have protected
conformal dimensions is not true without qualification.

What happens when the condition $J,J'\ll \sqrt{N}$ is not satisfied?
On the field theory side, one needs to go beyond the planar
approximation. Moreover, single-trace operators are no longer
orthogonal to multi-trace operators. On the string theory side, 
provided $g_s \ll 1$, single-string states remain orthogonal to
multi-string states. However, the description in terms of elementary 
strings is likely to be inadequate. This is indeed the case 
for states with $J'=0$, for which the correct semiclassical
description is known to be in terms of non-perturbative, rotating, 
spherical D3-branes, the so-called `giant gravitons' \cite{giant}. 
The operators dual to these states are not single-trace operators, but
(sub)determinant operators \cite{determinant}; the latter {\it are}
approximately orthogonal to each other if $J$ is comparable to $N$,
and only those with $J \leq N$ are independent from each other.   
A similar situation will presumably hold when both $J$ and $J'$
are non-zero. If this is the case, then the fact that the $\calo$-type 
operators are only independent if $2J'+J \leq N$ (since otherwise
they can be expressed as sums of products of operators of the same
type) will be irrelevant to the comparison with string theory, since 
these operators will only provide an accurate dual description of the
corresponding string theory states if $J,J'\ll \sqrt{N}$.

\sect{Discussion}
\label{discu}

Quantitative tests of the AdS/CFT conjecture that go beyond kinematics
are rare because a weak-coupling computation on one side generally 
corresponds to an strong-coupling computation on the other side. 
An exception to this state of affairs occurs in the sector of 
the rotating strings considered here, for two reasons 
\cite{FT03a,FT03b,FT03c,FT02}. First, the energy of the corresponding 
{\it classical} string configurations happens to admit an expansion 
in {\it positive} powers of $\l/(J+2J')^2$. Second, partial 
cancellations of sigma-model quantum corrections imply that all such 
corrections containing non-positive powers of $\l$ are suppressed in 
the limit $J + 2J' \gg 1$. These two facts allow the comparison of the 
string calculation to a perturbative SYM calculation in the regime 
in which $J+2J' \gg 1, \sqrt{\l}$. 

If $J'=0$ the strings considered here reduce to the BMN strings, that is, 
to point-like strings orbiting the $S^5$ around an equator, with angular 
momentum $J$ \cite{bmn}. The dual BMN operators are near-BPS operators, 
in the sense that they are `close to' (that is, `a few impurities 
away from') an exactly 1/2-supersymmetric operator, the so-called 
BMN ground-state; thus, in the BMN case, the agreement tests the 
AdS/CFT conjecture in a near-supersymmetric sector, and this fact is 
presumably responsible for the partial cancellations of sigma-model 
quantum corrections that are essential for the comparison to be possible.

It has not been appreciated previously that the situation is very
similar for the rotating strings discussed here with $J'\neq 0$. 
This is implied by the results of this paper, since we have shown that 
these strings asymptotically become 1/4- or 1/8-supersymmetric in the
limit of large angular momenta. 
A subtle difference between the extended strings and the BMN collapsed
strings case is, however, that the operators dual to the strings 
with $J' \neq 0$ are not near-BPS\footnote{
Except if $J' \ll J$, in which case they are BMN operators.} 
but {\it nearly}-BPS, in the sense that there is no
exactly 1/4- or 1/8-BPS operator that these operators are close to.

We would like to emphasize that tensionless strings arise in our 
analysis via an ultra-relativistic approximation in which the energy 
due to the string tension is negligible compared to the kinetic energy. 
In this sense, we think of the limit $\lambda/J^2 \ra 0$ as a limit 
in which $\lambda$ is kept fixed and $J$ is sent to infinity; this
is particularly natural in view of the fact that we also need 
$J \gg 1$. However, one can equivalently think of this limit as 
fixing $J$ to be much larger than unity and then sending 
$\lambda \ra 0$. In the strict limit $\lambda =0$ the rotating strings 
become exactly supersymmetric, and this may be relevant to the 
correspondence between tensionless strings in 
$AdS_5\times S^5$ and free $\caln=4$ SYM theory (see, for example, 
\cite{BMS03}). Note that the free field theory has an infinite
number of global symmetries, which could correspond to the gauge
symmetries of massless particles of all spin in the tensionless string
spectrum.\footnote{Tensionless strings could also arise as collapsed D3-branes, 
for example, but such configurations could probably not be justified 
within a semiclassical approximation.}

In this paper we have focused on {\it circular} rotating strings, 
but the same type of agreement between string energies and conformal 
field theory anomalous dimensions has been found
for other types of rotating strings, such as folded strings rotating 
in $S^5$ \cite{FT03c} and strings carrying angular momenta both in 
$S^5$ and in $AdS_5$ \cite{BFST03}. The natural question of whether 
these other types of strings also become supersymmetric in the limit 
of large angular momenta therefore arises. We hope to report on this 
in the future.

\section{Acknowledgments}

It is a pleasure to thank Eric D'Hoker, Dan
Freedman, Joaquim Gomis,  Paul Howe, Stefano Kovacs, Lluis Masanes,
Krystof Pilch,  Jorge Russo, Anton Ryzhov and Arkady Tseytlin for 
helpful discussions, and the Benasque Centre for Science for 
hospitality during the last stages of this work. PKT thanks the ECM
members of the University of Barcelona for hospitality, and ICREA for
financial support. 



\begin{thebibliography}{80} 


\bibitem{Maldacena97}
J.~M.~Maldacena,
``The large N limit of superconformal field theories and supergravity,''
Adv.\ Theor.\ Math.\ Phys.\  {\bf 2} (1998) 231
[Int.\ J.\ Theor.\ Phys.\  {\bf 38} (1999) 1113]
[arXiv:hep-th/9711200].

\bibitem{GKP98}
S.~S.~Gubser, I.~R.~Klebanov and A.~M.~Polyakov,
``Gauge theory correlators from non-critical string theory,''
Phys.\ Lett.\ B {\bf 428} (1998) 105
[arXiv:hep-th/9802109].

\bibitem{Witten98}
E.~Witten,
``Anti-de Sitter space and holography,''
Adv.\ Theor.\ Math.\ Phys.\  {\bf 2} (1998) 253
[arXiv:hep-th/9802150].

\bibitem{GKP02}
S.~S.~Gubser, I.~R.~Klebanov and A.~M.~Polyakov,
``A semi-classical limit of the gauge/string correspondence,''
Nucl.\ Phys.\ B {\bf 636} (2002) 99
[arXiv:hep-th/0204051].

\bibitem{FT03a}
S.~Frolov and A.~A.~Tseytlin,
``Multi-spin string solutions in AdS(5) x S**5,''
Nucl.\ Phys.\ B {\bf 668} (2003) 77
[arXiv:hep-th/0304255].

\bibitem{FT03b}
S.~Frolov and A.~A.~Tseytlin,
``Quantizing three-spin string solution in AdS(5) x S**5,''
JHEP {\bf 0307} (2003) 016
[arXiv:hep-th/0306130].

\bibitem{FT03c}
S.~Frolov and A.~A.~Tseytlin,
``Rotating string solutions: AdS/CFT duality in non-supersymmetric
sectors,'' Phys.\ Lett.\ B {\bf 570} (2003) 96
[arXiv:hep-th/0306143].

\bibitem{hoppe}
J.~Hoppe and H.~Nicolai,
``Relativistic Minimal Surfaces,''
Phys.\ Lett.\ B {\bf 196} (1987) 451.

\bibitem{stable}
P.~Demkin,
``On the stability of p-brane,''
Class.\ Quant.\ Grav.\  {\bf 12} (1995) 289
[arXiv:hep-th/9412172].

\bibitem{BMSZ03}
N.~Beisert, J.~A.~Minahan, M.~Staudacher and K.~Zarembo,
``Stringing spins and spinning strings,''
JHEP {\bf 0309} (2003) 010
[arXiv:hep-th/0306139].


\bibitem{AFRT03}
G.~Arutyunov, S.~Frolov, J.~Russo and A.~A.~Tseytlin,
``Spinning strings in AdS(5) x S**5 and integrable systems,''
arXiv:hep-th/0307191.

\bibitem{superconformal}
I.~Bandos, E.~Ivanov, J.~Lukierski and D.~Sorokin,
``On the superconformal flatness of AdS superspaces,''
JHEP {\bf 0206} (2002) 040
[arXiv:hep-th/0205104].

\bibitem{bmn}
D.~Berenstein, J.~M.~Maldacena and H.~Nastase,
``Strings in flat space and pp waves from N = 4 super Yang Mills,''
JHEP {\bf 0204} (2002) 013
[arXiv:hep-th/0202021].

\bibitem{AFSZ99}
L.~Andrianopoli, S.~Ferrara, E.~Sokatchev and B.~Zupnik,
``Shortening of primary operators in N-extended SCFT(4) and  harmonic-superspace analyticity,''
Adv.\ Theor.\ Math.\ Phys.\  {\bf 3} (1999) 1149
[arXiv:hep-th/9912007].

\bibitem{Ryzhov01}
A.~V.~Ryzhov,
``Quarter BPS operators in N = 4 SYM,''
JHEP {\bf 0111} (2001) 046
[arXiv:hep-th/0109064].

\bibitem{DDHR03}
E.~D'Hoker, P.~Heslop, P.~Howe and A.~V.~Ryzhov,
``Systematics of quarter BPS operators in N = 4 SYM,''
JHEP {\bf 0304} (2003) 038
[arXiv:hep-th/0301104].

\bibitem{DO03}
F.~A.~Dolan and H.~Osborn, ``On short and semi-short representations for
four-dimensional superconformal symmetry'', Annals Phys.
{\bf 307} (2003) 41 
[arXiv:hep-th/0209056]

\bibitem{HH03}
P.~J.~Heslop and P.~S.~Howe,
``Aspects of N = 4 SYM,''
arXiv:hep-th/0307210.

\bibitem{giant}
J.~McGreevy, L.~Susskind and N.~Toumbas,
``Invasion of the giant gravitons from anti-de Sitter space,''
JHEP {\bf 0006} (2000) 008
[arXiv:hep-th/0003075].

\bibitem{determinant}
V.~Balasubramanian, M.~Berkooz, A.~Naqvi and M.~J.~Strassler,
``Giant gravitons in conformal field theory,''
JHEP {\bf 0204} (2002) 034
[arXiv:hep-th/0107119].

\bibitem{FT02}
S.~Frolov and A.~A.~Tseytlin,
``Semiclassical quantization of rotating superstring in AdS(5) x S(5),''
JHEP {\bf 0206} (2002) 007
[arXiv:hep-th/0204226].

\bibitem{BMS03}
M.\ Bianchi, J.\ F.\ Morales and H.\ Samtleben,
`` On stringy $AdS_5 x S^5$ and higher spin holography'',
JHEP {\bf 0307} (2003) 062
[arXiv:hep-th/0305052].

\bibitem{BFST03}
N.~Beisert, S.~Frolov, M.~Staudacher and A.~A.~Tseytlin,
``Precision spectroscopy of AdS/CFT,'' arXiv:hep-th/0308117.



\end{thebibliography}
\end{document}